# Quantifying the Need for Attorney Pro Bono Services in Connection with the Social Determinants of Health

January 7, 2021

Yi Mao, MPH -- Editor
Stacey R. Beck, MS -- Analytics and Visualizations

Contributors:
Benjamin Bartek, JD
Beatriz Cabrera
Rachell Calhoun
David Coe
Jakob Cronberg
Suren Nalluri
Bradley Merrill Thompson, JD, MBA

The authors are students in the Master of Applied Data Science program at the University of Michigan School of Information. This paper summarizes a required milestone project for the program.

Acknowledgements

We want to thank deeply the instructors for the course for their feedback and guidance:

Christopher Teplovs, PhD
Anthony Whyte, MA
Yumou Wei, MS

We also want to thank Sherri Dixon, PhD, Senior Biostatistician at National Center for Healthy Housing, for her guidance on understanding available housing data. We also want to thank Chase Haller, Neighborhood Christian Legal Clinic, for all of his help in creating appropriate weights for the factors that determine whether an apartment is legally substandard and more generally for the education he gave us on the role of the legal aid attorney in housing matters.

CONTENTS

## I. Introduction

The paper estimates the need for additional attorney hours annually to address the legal needs of indigent clients throughout the United States in matters that comprise the so-called social determinants of health (SDoH). The result will inform stakeholders such as policy makers and private donors so they can allocate resources appropriately and design programs to close the "justice gap." As a pilot study, the scope of the project covers only a few major justice problems related to the social determinants of health (standard housing, evictions and foreclosures, guardianships for the incapacitated, and victims of domestic violence) because they significantly impact health outcomes and there are data available.

## II. Summary

Based on our calculations, we estimate that the total number of attorney hours to address only these five legal issues is over 34 million per year. To put that in perspective, the American Bar Association estimated that in 2018, the year from which much of our data comes, the total number of practicing attorneys in the United States was 1,338,678 (Weiss). Thus, to provide the needed hours, every single practicing attorney in the United States would need to contribute about 26 hours a year. While many lawyers do in fact contribute pro bono hours, they address the full range of legal needs that go well beyond just the five we studied.

## III. Context

The U.S. Pledge of Allegiance represents that all individuals should have justice. However, there has been growing awareness of a "justice gap" in the U.S., particularly among the low-income group. "Justice Gap" refers to the gap between civil legal needs and resources available. Reportedly, in 2017, 86% civil legal problems received either insufficient or no legal help within the low-income population (The Legal Services Corporation, 2017).

Evaluating the need for pro bono services may help policy makers design programs to close the gap. Project Protect attempts to build a model that estimates the need for pro bono services. The goal of Project Protect is to identify how many people who need legal services also live in poverty and therefore require pro bono legal aid.

As a pilot study, Project Protect focuses on justice issues that have implications for the social determinants of health. The project assesses the need for pro bono services in terms of four areas including substandard housing, evictions and foreclosures, guardianships for the incapacitated, and victims of domestic violence. We adopt this focus for three reasons. First, these are among the most compelling legal needs that low income Americans have. Second, they are among the most common legal needs that low income Americans have (The Legal Services Corporation, 2017). Third, these issues are key SDoH that impact health outcomes (DPHP, 2020).

### A. The Need for Legal Aid Generally

Americans do not have the guarantee of an attorney for civil matters. The Constitution only guarantees an attorney for criminal matters. Unfortunately, the current system leaves unattended legal needs among those who cannot afford legal assistance for civil matters.

Although there are, of course, legal aid clinics, and some attorneys choose to offer free services pro bono, those existing resources cannot fully meet the needs. Legal aid clinics are woefully understaffed, but try to help people who have an income typically under 150% of the Federal Poverty Guidelines. In 2017, reportedly 86% civil legal problems received either insufficient or no legal help within the low-income population (The Legal Services Corporation, 2017). More specifically, 71% of low-income households need legal assistance for at least one civil legal problems with housing condition, domestic violence, health care, and disability access (The Legal Services Corporation, 2017). The National Center for Access to Justice (NCAJ) uses data to analyze what resources are available for civil legal aid. The center conducts a periodic survey in order to quantify those resources. The center launched an update survey in 2020, but because of COVID, collecting and publishing the results has been delayed (NCAJ, 2020).

### B. Social Determinants of Health (SDoH)

SDoH is defined as "the conditions in which people are born, grow, live, work and age, and the wider set of forces and systems shaping the conditions of daily life" by World Health Organization (WHO). There is compelling evidence from literature showing that SDoH account for 30-55% of health outcomes (WHO, n.d.). The SDoH are further defined as societal circumstances that directly or indirectly affect a person's health like:

(1) the quality of a person's housing,
(2) their mental health status and whether they are protected or not,
(3) whether they are getting adequate public financial support for food and healthcare, and
(4) whether they are being victimized by crime, for example, domestic violence.

Addressing the SDoH needs is critical to improve population health and promote equity in health. Pro bono service is one of the fundamental resources required to close the gap of SDoH needs. Further, from the perspective of data availability and accuracy, there are many public resources of data tracking health and the social determinants. For instance, CDC's data sources for SDoH allows for our analysis (Sources for Data on SDOH).

## IV. Current Estimation Methodologies

Here are five of the more common methodologies currently used to estimate the unfulfilled need for legal aid in America.

### A. Extrapolation from the Utilization of Attorneys by Those Who Are Not Indigent

Most of the present estimates for the size of the justice gap between the need for legal aid and legal aid provided uses as a methodology a determination of the legal services obtained by those who are not indigent, and then extrapolates to who are. For example, if those above the poverty line have access to attorneys at a certain rate, we extrapolate that figure to those below the poverty line to assess whether there are sufficient legal aid attorneys to serve them. An example of this methodology is the computation of the Justice Index (Attorney Access Methodology).

For example, one group estimates that nationally, on average, only one legal aid attorney is available for every 6,415 low-income people. By comparison, there is one private attorney providing personal legal services for every 429 people in the general population who are above the poverty threshold (Legal Services Corporation, 2009).

The problem with that is the nature and volume of the need for legal services among those living in poverty should not be considered necessarily analogous to the nature and volume of need for legal services desired among those living above the poverty line. The two groups face different legal issues. Those above the poverty line, for example, might utilize attorneys more for estate planning because they have more financial resources to divide upon their deaths. Those below the poverty line may have more need for attorneys to address things like substandard housing and advice on government benefits.

### B. Surveying those Living in Poverty About Their Legal Needs

A second approach utilized to estimate the need for attorneys among the poor is to survey the poor. For example, the Legal Services Corporation (LSC) contracted with NORC at the University of Chicago to help measure the justice gap among low-income Americans in 2017. NORC conducted a survey of approximately 2,000 adults living in households at or below 125% of the Federal Poverty Level (FPL) using its nationally representative, probability-based AmeriSpeak® Panel. This report presents findings based on this survey and additional data LSC collected from the legal aid organizations it funds (Attorney Access Methodology). A justice gap study done in Wisconsin used the approach as well (Bridging the Justice Gap).

There are a couple of challenges with utilizing survey data in this instance. It is difficult to ask meaningful survey questions about current legal needs in a survey delivered to a lay population.Written questions may be misunderstood, and oral interviews inject the risk of bias.

### C. Tracking the Number of Clients Who Are Turned Down for Legal Aid

This approach is rather directly relevant, but logistically difficult. This was calculated based on a study where legal aid attorneys around the country in one organization kept records on the number of people who they turned down, and provide that data to a central data repository (Documenting the Justice Gap In America).

Word gets around quickly. If legal aid groups are turning people away because they do not have the resources to help them, there may be many people who never seek legal aid to begin with because they think it likely they will be turned down. Further, it is just one organization and it is hard to extrapolate to all legal aid based on those data. In addition, if the concern is that there are not attorneys present to be consulted, for example in rural areas, then there is no instance where a client would be turned down.

### D. Tracking the Number of Litigants in America Who Represent Themselves

While this approach is concrete in that it identifies the number of laypeople who try to represent themselves in litigation without an attorney (Documenting the Justice Gap In America), it has two obvious limitations. First, it only considers litigation and not the myriad of other legal needs that people have. Second, it does not reflect the number of Americans who are daunted by the task of representing themselves, and therefore choose not to.

### E. Combination of Quantitative and Qualitative Assessments

Some groups simply choose to estimate the number based on a wide variety of assessments, and then compiling all the data together using a variety of methodologies. For example, a New Hampshire study employed a multi-method approach that included interviews with social services providers, questionnaires of volunteer attorneys, analysis of legal services data, a lengthy survey of judges, clerks, and administrators in the New Hampshire system, and analysis of legal needs assessments from other states and public data sources such as the U.S. Census (Study of the Legal Needs of New Hampshire's Low-Income Residents).

The bottom line is that all of these approaches offer their own strengths but also suffer from one or more limitations. Thus, we wanted to see whether there might be a new approach that would either reinforce existing estimates or produce different results.

## V. Methodology

This project will use a new methodology for this estimation that hinges on:

1. An estimation of the magnitude of the SDoH that legal aid attorneys address using published data regarding the scope of those problems;
2. Collected survey data from practicing legal aid attorneys to estimate the average utilization of attorney time per issue;
3. Combining the results of 1 and 2 to estimate the number of attorney hours across the country –at a regional level – required to address those SDoH issues.

We were able to tackle four such social determinants, namely substandard housing, evictions and foreclosures, guardianships and assisting victims of domestic violence.

## VI. Survey – Attorney Resource Utilization in Addressing the Social Determinants

### A. Design and Distribution

Throughout the summer of 2020, we distributed a survey to practicing attorneys asking them how long it takes, on average, to assist a client in the areas that interested us. We included ten different matters because at that time, we were not certain how many of the SDoHs we could analyze. The distribution was through a Google form, and we encouraged attorneys to participate through social media posts, primarily LinkedIn. We also enlisted several national organizations with attorney members to encourage their members to participate. To encourage participation, on the basis of a donation we received, we offer to give $5000 to the legal aid clinic of choice for a random winner selected from all those who participate.

Before asking the typical hours spent on each type of case, we collected information on the state and ZIP Code of the respondent. We also asked the respondents to choose between: Full Time--I am a legal aid professional, and Part Time--I do other work primarily. This is an important factor, because those who do legal aid professionally are likely to be more efficient at handling the cases than a volunteer who only does the work occasionally. That said, because our legal aid system relies on both kinds of service, we wanted to make sure we got a balance between the two.

In answering the questions, we prefaced the survey with the following instructions: "Please fill in the estimated total time in hours, on average, that it takes you to handle a case in the indicated matter. Omit any areas where you do not feel you have adequate experience. As a minimum, please do not answer questions for any area where you have not handled at least three such cases in the last 3 years."

## B. Findings

470 attorneys completed the survey. Those attorneys were distributed across the United States, with the largest number of participants coming from Texas, Illinois and Florida. With regard to occupations, 339 respondents were full-time legal aid attorneys, while 127 were volunteers. One person filled in the other category, which we are honestly not sure what that is. The top three specializations in terms of number of respondents are domestic abuse, eviction, and substandard housing. In terms of the average estimated hours per case, the top three time-consuming categories are mortgage foreclosures, substandard housing, and domestic abuse.

|  | Count of respondents | Median hours per case | Average hours per case |
| --- | --- | --- | --- |
| Substandard housing | 174 | 10.0 | 27.54 |
| Mortgage foreclosures | 112 | 25.5 | 72.59 |
| Eviction | 223 | 10.0 | 23.92 |
| Domestic Abuse | 241 | 15.0 | 26.55 |
| Guardianship | 122 | 10.0 | 19.09 |

## C. Limitations and Assumptions

In this analysis, we are using the median score instead of the mean score because we think the median is more likely to be accurate. We observed what we think were two types of errors. First, some people filled in their answers with regard to certain matters with a zero, which we assume was probably meant to convey that they have insufficient experience on that particular kind of case to respond. They should have simply not filled that in, so if they put in a zero, we removed it because we do not believe that anyone can handle a legal case in literally zero hours.

There were others who listed the average hours in the hundreds. We think they probably meant to write, for example 3000 to mean 30 hours and 00 minutes. In any case, it was only a couple of answers, but they greatly affected the mean. Thus, we chose the median – or middle representation – as more accurate.

Further, data collection was biased in the following sense. There were leaders in legal aid groups who were willing to help us get the word out, and they did so with differing effectiveness. For example, a senior leader in Illinois sent a blast email out to all of her contacts in Illinois, which produced over the next few days a large influx of Illinois responses. In other states, such as California, we did not have a similar advocate so we have less representation from California. In other words, responses were driven by how well connected we were in a state. That said, there were several national organizations that encourage people nationally to participate in so we do have data from 33 states. We believe that is important, because state laws vary, and so the time that an attorney might have to spend in different states could be expected to differ significantly.

We did not ask survey participants to go back through their accounting data to calculate actual averages. Thus, these averages are based simply on subjective assessments. We recognize that some survey respondents might wish to suggest that they are efficient in what they do, and thus estimate low. We further recognize that other survey respondents might wish to suggest that more resources are needed, and estimate high. We are comforted to see that there is a natural bell curve for most of these data, suggesting that there is at least some consistency and the average seems reasonable.

## VII. Findings

### A. Substandard Housing

#### 1. Issue

Many people in America live in rental units that constitute substandard housing. Even if the unit violates a local housing code or local public health codes, the tenant often finds the landlord resistant to making the needed improvements. Many landlords simply do not want to invest in the necessary repairs, and often they perceive people living in poverty as without the power to force them to do anything. Because the conditions violate the local housing code, the local health code or the terms of the lease, an attorney can assert a claim against the landlord. Often these claims are settled out of court, but it takes the involvement of an attorney to motivate the landlord.

#### 2. Data Used and Time Period Analyzed

We used the American Housing Survey from the years 2017, 2015, and 2013 created by the Department of Housing and Urban Development and contained in census data because these versions of the survey include very specific and pointed questions about the condition of homes in which people dwell, and includes different metro areas for each of those years surveyed. Every other year, for this housing data the census surveys a group of the largest metropolitan areas, and then on a rotating basis surveys smaller metropolitan areas. Therefore, by combining multiple years, we obtained data on 119 different metropolitan locations.

#### 3. Method

##### (a) Estimating Total Cases Where a Lawyer Would Be Useful

For each metropolitan area, we calculated a percentage of homes surveyed that were substandard. To discern which surveyed apartments were likely out of compliance with building codes, we consulted an experienced legal aid housing expert who helped us construct an elaborate weighting system for the relevant factors collected in the American Housing Survey. The weights reflect his judgment about the meaning of the survey results in terms of whether an attorney would likely conclude that the apartment dweller has a claim against the landlord for substandard housing.

To estimate the total number of substandard apartments in a given region, we multiplied that survey percentage by the total number of apartments in the metropolitan areas. We obtained the total number of apartments from the 2018 census data.

(b)   Applying the Poverty Filter

The American Housing Survey includes data on income levels. To determine how many apartment dwellers would qualify for legal aid, we applied the Federal Poverty Guidelines for the appropriate year.

4.   Nationwide Map Showing Need

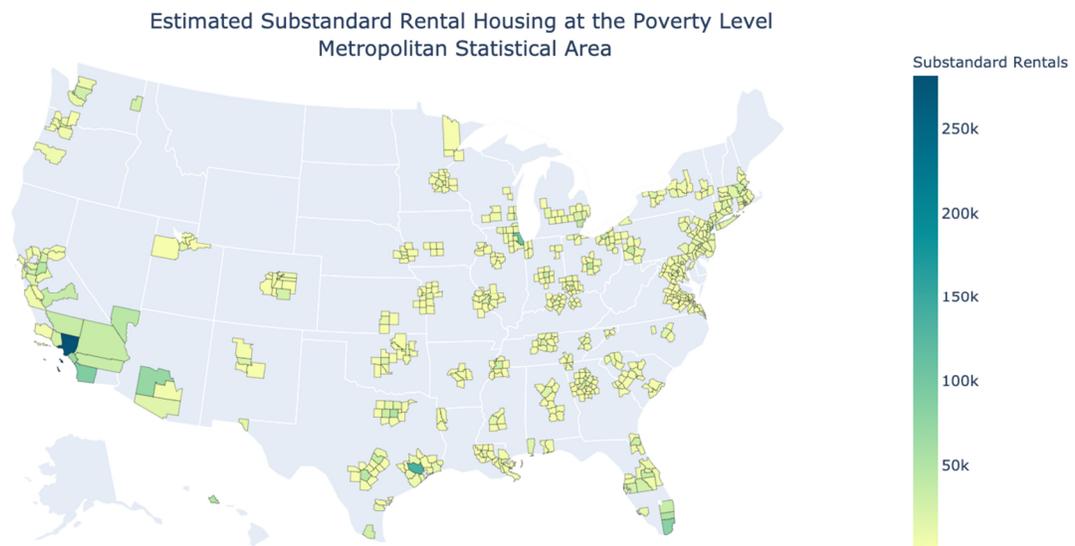

5.   Total National Number of Cases

Based on our data analysis, we estimate that 4,840,000 households in 119 large metropolitan areas across United States have household incomes under 150% of the Federal Poverty Guidelines and live in apartments under such conditions that the household could assert a claim that the apartment is unfit or in violation of local building codes. These households therefore would benefit from having access to a legal aid attorney to help them improve their living conditions.

6.   Assumptions and Limitations of Analysis

- Some landlords will voluntarily improve the conditions of a rental unit without the involvement of an attorney. Unfortunately, this is unusual in the situation where an apartment has deteriorated to the point that it violates local housing or public health codes. In our analysis, we assume that some involvement by an attorney in each case of substandard housing will be beneficial.

- The housing analysis is significantly limited to only the 119 metropolitan areas included in the last three years of AHS surveys. This leaves out not only the entire rural areas of the United States, but it also leaves out many significant but slightly smaller cities and college towns across the country. The analysis excludes 34.65 % of the US population.
- The weights assigned are the opinions of two experts. Mr. Haller is an expert with considerable relevant experience, but there may well be a diversity of views. The census organization did not draft the AHS questions with compliance with legal building codes in mind, so in many cases the questions were written imprecisely for our purposes. We thus had to apply subjective judgment in several cases.
- We would note that our analysis is not overly sensitive to the weights, in that apartments ended up counting as substandard as long as there were a total of at least three relevant problems for a given apartment unit. Thus, the main issue is whether we accurately identified the lower end of the weight scale. As a matter of quality control, we closely looked at the impact of the weights on the findings of individual surveys, and found that no individual variable accounted for more than 10% of the violations, and the vast majority were well below that. Consequently, if there is debate about a single variable, a single variable would not significantly impact the findings.

### B. Evictions and Foreclosures

#### 1. Issue

Foreclosure and eviction proceedings are complicated proceedings where a defendant benefits considerably by having an attorney. In both types of proceedings, there are many rights that the homeowner or renter has that are not well-known. Thus, we assumed that in each case of foreclosure or eviction, the defendant would benefit from having an attorney to advise them.

#### 2. Data Used and Time Period Analyzed

For evictions, we used nationwide eviction and poverty data from Eviction Lab for the years 2000 through 2016. The Eviction Lab is a team of researchers at Princeton University who believe that a stable, affordable home is central to human flourishing and economic mobility. The Eviction Lab has published the dataset of evictions in America, going back to 2000.

Publicly available nationwide foreclosure data in general is sparse. However, in response to the housing crisis in 2008, the U.S. Department of Housing and Urban Development compiled data estimating the number of foreclosures across the United States for the period of January 2007 – June 2008 at the state and county level for the purpose of allocating Neighborhood Stabilization Program grants. While it is not the only dataset from which mortgage information can be extrapolated, it is a comparatively recent, comprehensive, nationwide, non-private compilation of foreclosure statistics. More foreclosures likely occurred in 2008 compared to a typical year, however, the data is informative and may prove particularly prescient in anticipating a possible surge in foreclosure cases as the current pandemic-driven economic crisis persists.

3. Method

(a) Evictions

The choropleth map below depicts the eviction rates in the country over the years 2000 – 2016.

(b) Foreclosures

We calculated the total number of foreclosures for each of the individual states for the 18-month period from January 2007 – June 2008. We calculated a 12-month estimate for 2007-2008 by multiplying the 18-month totals by 2/3.

We then multiplied the number of foreclosures per year by the percentage of households living below the 150% poverty threshold in each state in order to achieve the desired estimate of foreclosure actions in which legal assistance is needed due to financial hardship. This calculation likely includes both well-established and indigent households, as well as middle class households who have fallen on financial hardship and may even fall below the 150% poverty threshold by the time a foreclosure action is filed.

4. Nationwide Map Showing Need

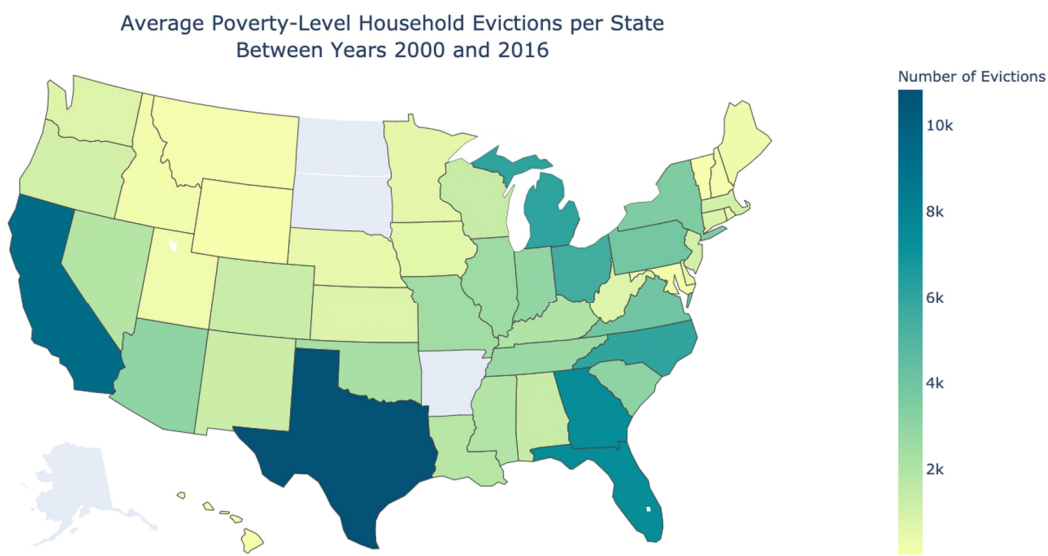

5. Total National Number of Cases

(a) Evictions

The mean nationwide eviction rate for the years 2000-2016 was 2.81%. After applying the poverty rate state-by-state against the number of evictions, the number of evictions nationwide for those living in poverty is 105,440.

(b) Foreclosures

In all, 2,922,063 foreclosure actions were initiated during the 18-month period from January 2007 – June 2008. Reduced to a 12-month period, an estimated 1,948,042 foreclosures actions were initiated across the United States. Nationwide, when combined with the 150% poverty threshold data for each state, an estimated 440,299 foreclosure actions required legal assistance over the 2007-2008 12-month time period.

6. Assumptions and Limitations of Analysis

Our analysis is limited to the time periods and geographic areas specified in the report.

(a) Evictions

- For our eviction data, our analysis relies on the accuracy of the underlying data collected by The Eviction Lab and government agencies. We understand from the literature that there are some limitations on the quality of the data from The Eviction Lab.

- To calculate eviction rates, we made two balancing assumptions. The first assumption is that the people being evicted reflect the same income demographics characteristic of the general population. In other words, we assumed that a person being evicted was no more likely to be living below the poverty level than any member of the general population. We think that assumption is clearly wrong, in the sense that people being evicted are, we suspect, much more likely to be living in poverty than those not being evicted. However, we do not have income data on those being evicted.

- The other assumption we made is that everyone who is poor and being evicted could benefit from using an attorney. It is important to understand that in eviction proceedings, they typically do not focus simply on whether the rent was paid. Eviction proceedings are designed to produce a just outcome that ensures the person being evicted has adequate notice of the eviction and adequate time to find alternative living quarters. We consulted an expert in the field, Chase Haller, who supported this assumption. He explained that almost always in an eviction he can add value for the tenant by negotiating an orderly departure. Sometimes this includes a reasonable move out date that avoids an eviction judgment. Avoiding an eviction judgment is useful to the tenant because of the impact an eviction has on credit score. Sometimes the attorney is able to negotiate a payment plan that allows the tenant to remain in the apartment. Indeed, professional legal aid attorneys are often aware of rental assistance programs, for example in this time of the COVID pandemic, that allow the tenant to make up the delinquencies. Often the tenant has counterclaims against the landlord that can be used in a negotiation to assure an equitable result. Evictions also sometimes include issues of what is to be done with the tenant's personal property. Thus, it is very often useful to have an attorney to ensure that the eviction process is fair, even if nonpayment of rent factually is not at issue.

- We believe that those two assumptions in some measure balance each other out, but still produce a very conservative estimate. We think it is beyond doubt that people being evicted are more likely to be living in poverty than the general population. At the same time, we also suspect that some-- but honestly very few – evictions are straightforward enough that having access to a lawyer would not provide any particular benefit to the tenant. To be clear, in the experience of the lawyers involved in this project and the experts we consulted, that is unusual because landlords aggressively pursue their interests and those who are impoverished lack the power to resist.

    (b)  Foreclosures

- For foreclosures, our theory was that many people who buy a home, at the time they buy the home, are likely not living in poverty. While foreclosure indicates a turn of circumstances, that does not tell us whether the person is actually now living in poverty. Someone might buy an expensive home in good times, and then not be able to make large payments in bad times. That does not mean they are living in poverty. Thus, similar to evictions, the safest, albeit conservative assumption, was to simply apply the percentage living in poverty in that state to the population of people whose homes are being foreclosed. We think this is likely understating the number of people in foreclosure who, at the time of foreclosure, live below the 150% Federal Poverty Guidelines limit.
- We also made the assumption that everyone in foreclosure could benefit from hiring an attorney. We believe this assumption is even more reasonable than with evictions, because foreclosures are indeed much more complex procedurally and the issue is not simply whether the person ought to leave the premises. The financial implications of a foreclosure are much more significant, as between the parties and with respect to the borrower in particular.
- On the other hand, we are aware that the time period we selected for studying foreclosures was one of the worst in recent years. Thus, while the time period would suggest overstatement of the final numbers, the income analysis would suggest understatement, to some extent balancing each other.

  **C.**  **Guardianships for the Incapacitated**

    1.  Issue

Cognitively disabled individuals living in poverty often need increased institutional support, such as a legal or financial guardian, to allow for equal access to opportunities and legal protections, such as decreasing the risk of financial exploitation by an unethical employer.

Unfortunately, the process of establishing a guardian is in itself a legal procedure that has significant barriers, including cost. As such, the goal of our project is to estimate how many people in different geographic areas may be living with a disability that requires them to interact with the legal system, while also lacking the funds to hire an attorney to aid in establishing guardianship.

In this analysis, we generally refer to the need for guardianship. We do, though, fully recognize that many people may not need a guardianship per se, but may need other more appropriate legal help. We understand that different states offer different alternatives for protecting those with cognitive disabilities. For ease of reference, even though it is not legally precise and correct, we refer to all legal vehicles for helping the cognitively disabled as guardianships. Our point is simply to calculate the number of people who need legal aid on the basis of a cognitive disability, regardless of exactly the purpose of that legal aid.

2. Data Used and Time Period Analyzed

Every year, the U.S. Census Bureau contacts over 3.5 million households across the country to participate in the American Community Survey (ACS) by sampling 250,000 households per month and producing estimates of various demographic features for geographies with populations of 65,000 or greater. For this project, we used the ACS yearly population estimates of civilian, noninstitutionalized populations for whom poverty status was determined, spanning from 2014 to 2018, and concentrating on metropolitan areas. For county-level population estimates, we used a dataset downloaded from the Open Science Foundation, with estimates calculated based on historical census data from 1990 to 2015.

3. Method

(a) Estimating Total Cases Where a Lawyer Would Be Useful

Our goal was to calculate the need for legal aid in situations where the individual would qualify for guardianship, but also lack the financial resources to obtain it themselves.

Seeking guardianship can be intensely personal decision, and in recognition of the need to honor an individual's right to be an active participant in their legal and medical status, our project has decided to operate under the assumption that individuals that self-report having difficulty with cognitive tasks on the American Community Survey would be likely candidates for guardianship. Although not an ideal indicator, there is currently a paucity of data examining best practices for determining guardianship need at this time, particularly in regards to the intersection of cognitively disabled and impoverished individuals.

The relevant ACS question reads as follows: "Because of a physical, mental, or emotional condition lasting 6 months or more, the person has difficulty learning, remembering or concentrating."

(b) Applying the Poverty Filter

The American Community Survey allows for filtering by ratio of income to Federal Poverty Level. We were therefore able to calculate population estimates of people with a self-reported cognitive disability having incomes at or below 150% the Federal Poverty level for each metropolitan area.

We established a growth moving average based on the 2014-2018 American Community Survey data, which we then merged with the county level population estimates to calculate an estimate for our population of interest for 2018 by county.

4. Nationwide Map Showing Need

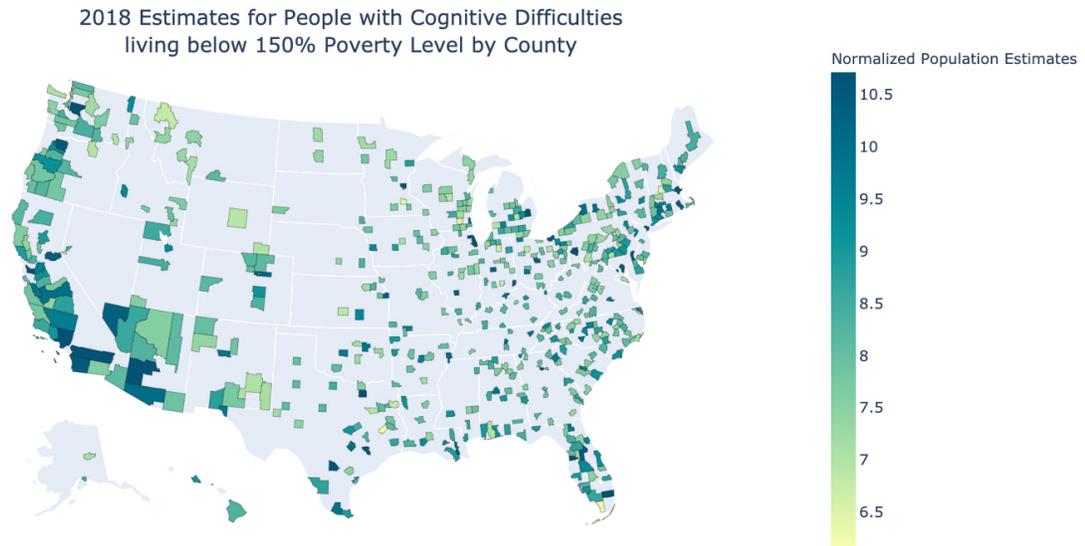

5. Total National Number of Cases

Based on our data analysis, we estimate that 4,369,183 individuals in 511 counties with large metropolitan areas across the United States have incomes under 150% of the Federal Poverty Guidelines and self-report cognitive difficulties sustained over the past 6 months. These individuals would therefore likely benefit from having access to a legal aid attorney to help establish guardianship.

6. Assumptions and Limitations of Analysis

While our analysis is meant to be an initial foray into a topic of interest and in no way represents a conclusive report, it indicates a need for further investigation into this subject. Our estimate of 4.4 million individuals is likely a conservative estimate, as there are a number of features related to the population of interest that could lead to under-reporting:

- Vulnerable populations are more susceptible to housing insecurity, which could result in them not being counted in the ACS survey.

- Vulnerable populations may also be harder to contact, due to decreased access to technology and more frequent moves.

- People may choose not to self-report cognitive difficulties when asked for personal or healthcare reasons.

- Degenerative cognitive decline is often not recognized by individuals or their families in its initial stages.

- Additionally, individuals in our population of interest might struggle with completing the survey due to time, literacy, or other pressures.

- Finally, the US population living in rural communities or smaller cities that was not included in our data is about 13% of the total population.

- On the other hand, the data do not tell us whether those responding might already have a guardianship. It is reasonable to believe that some do.

Additionally, although we used a five-year growth average for each metropolitan area, some areas may have missing data for one or more years in a metro area, which could have skewed the results.

### D. Victims of Domestic Violence

#### 1. Issue

The COVID-19 pandemic has worsened the experience of those experiencing domestic violence. There is evidence to suggest that domestic violence increases during times of disaster (Enarson, 1999) and during times of recession (Medel-Herrero et al., 2018). So far, the 2020 data show that such a change might be happening during this pandemic. Compared to this same period in previous years, hospitals have seen more evidence of domestic violence (Gosangi et al., 2020) and domestic violence hotlines have received an increased number of calls (National Domestic Violence Hotline, 2020).

This already underreported form of violence has also become harder to estimate as intimate partners currently have fewer opportunities to call for help when lockdowns keep them confined with their abusers (Li & Schwartzapfel, 2020). It remains uncertain how the ongoing pandemic will affect domestic violence statistics. But we believe it is essential to at least estimate a baseline level of need that can be used to improve access to the legal help that people experiencing domestic violence need (such as obtaining restraining orders or safeguarding the assets they need to start a new life), now more than in recent history.

#### 2. Data Used and Time Period Analyzed

We base our estimates on a combination of data from the U.S. Census Bureau's 2018 Annual Estimates of the Resident Population and from the 2018 National Crime Victimization Survey (NCVS). The NCVS is conducted twice each year by the U.S. Census Bureau in coordination with the U.S. Department of Justice's Bureau of Justice Statistics. Because domestic violence is historically underreported, we chose the NCVS because a) it uses a nationally representative sample of U.S. households and b) it is one of the foremost sources of information on unreported crimes (Inter-university Consortium for Political and Social Research, 2020). We used the data from the 2018 survey because it was the most recent and most relevant.

#### 3. Method

##### (a) Filtering the Data For Domestic Violence

Because the NCVS has information about crimes other than just domestic violence incidents, we first filtered for only survey respondents who had experienced at least one incident where the type of crime and their relationship to the offender fit the standard definitions of domestic violence. We obtain the standard definitions from the National Institute of Justice (National Institute of Justice [NIJ], 2007), the Department of Justice (U.S. Department of Justice, n.d.), and most state legislatures according to the National Conference of State Legislatures (National Conference of State Legislatures, 2019). Then we applied another filter to narrow the list to only those whose households living below 150% of the 2018 Federal Poverty Guidelines.

(b) Estimating Percent of Domestic Violence Among NCVS Respondents

We grouped the results by U.S. region, and for each region we calculated the estimated percent of people having experienced domestic violence in the region out of the total population within that region who were represented in this survey. During this process, we followed the NCVS guidelines on how to properly adjust for the fact that survey respondents each represent different amounts of the U.S. population.

(c) Estimating Domestic Violence Totals Among U.S. Population

Finally, we multiplied the estimated regional percentages by the 2018 U.S. Census population total estimates for their respective regions in order to approximate the number of people experiencing domestic violence in each U.S. region during 2018.

4. Nationwide Map Showing Need

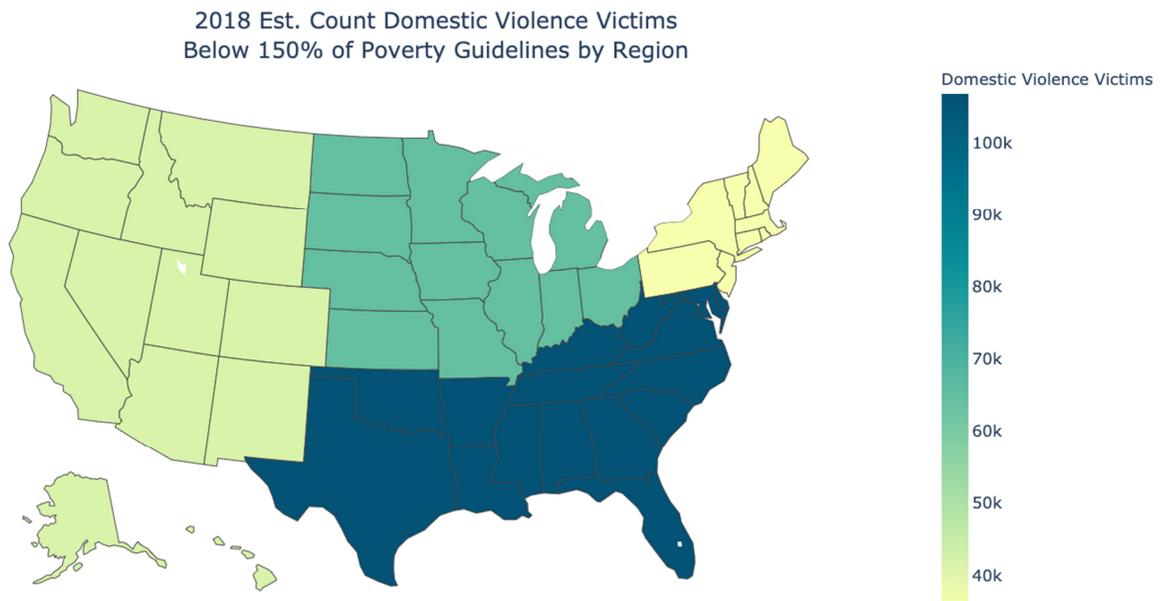

5. Total National Number Of Cases

The NCVS 2018 data combined with the 2018 Census population estimates tell us that in 2018 there were about 249,065 Americans who experienced domestic violence and who were also living below 150% of the Federal Poverty Guidelines.

6. Assumptions and Limitations of Analysis

- Because the NCVS location information is limited to regions. Hawaii and Alaska are in the West, but their poverty guidelines differ from the other 48 contiguous states and the District of Columbia (Assistant Secretary for Planning and Evaluation, 2018). If metropolitan area or state of residence data had been available, individuals living in Alaska or Hawaii would have been more accurately assessed for financial eligibility and our estimate for the West might have been different.

- The NCVS only includes household income information in the form of an income range. Our model measured each household's financial qualifications according to the upper limit of their given income range. More granular income information would have helped to assess more fairly each individual's financial qualifications. For example, in 2018 a household of two people would need to earn less than $24,690 in order to be below 150% of the Federal Poverty Guidelines (i.e., 150% of the sum of $12,140 for the first person and $4,320 for the second person). However, using our data model, if this household makes $20,000 (a qualifying amount), they are included in the range of people assumed to make $24,999 and are therefore ineligible. Had we used the lower bound of the income range, we would have encountered the opposite problem wherein some households that earn too much to qualify would be assumed to be eligible anyway.

- The National Institute of Justice includes stalking and psychological/emotional violence as forms of domestic violence (NIJ, 2007). Unfortunately, these forms of violence are beyond the scope of the NCVS and could not be captured in our analysis.

## VIII. Discussion

### A. Nationwide Total of Attorney Hours Needed Annually to Address the Need

1. Convert Each to an Annual Number

Two of the numbers we generated are snapshots in time. First, the estimate of the substandard housing is the number of housing units at any given time that might be substandard. Second, the number of people suffering from intellectual disabilities is similarly a snapshot in time. The goal of the project, however, is to estimate the annual investment society should make to ensure that those living in poverty are represented on these issues. It would be unrealistic to suggest that we need to solve all of these two problems in the first year.

Further, we did not quantify the incremental additions year to year, nor the incremental resolutions of these two issues year to year. For intellectual disabilities, new people are born every year that may, upon turning 18, need a guardianship. Likewise, people with disabilities die each year. Again, we simply took a snapshot at one point in time.

In some cases, a substantial investment of resources into attorney representation might also change the dynamic. It might be that if there were more attorneys willing to take cases on behalf of those with substandard housing, that more landlords might on their own improve living conditions. We can only speculate as to what might happen.

We decided that for the two areas where we are taking a snapshot, that in order to convert that data from a single point in time to an annual rate, we would divide by five. In other words, we felt it represented a reasonable degree of progress if we could take on 20% of the problem each year. We certainly would not anticipate that the problem would go away after five years. We just do not have a way to predict the future, and how it might be impacted by more attorneys working on a given issue. This is a pretty theoretical problem to worry about, because it is unlikely that we will achieve that level of attorney involvement anytime soon.

2. Multiply By Resource Utilization According to Survey

In the table below, we take the annual number of cases we want to tackle, we multiply it by the median number of hours that attorneys reported in the survey as necessary to tackle such cases, to produce an annual number of attorney hours needed to tackle those cases. We then summed each of those categories to produce the grand total.

|  | Median hours per case | Annual cases | Annual hours needed |
|---|---|---|---|
| Substandard housing | 10.0 | 968,000 | 9,680,000 |
| Mortgage foreclosures | 25.5 | 440,299 | 11,227,624 |
| Eviction | 10.0 | 105,440 | 1,054,400 |
| Domestic Abuse | 15.0 | 249,000 | 3,735,000 |
| Guardianship | 10.0 | 873,863 | 8,738,630 |
| Total |  |  | 34,435,654 |

B. **Accuracy of the Estimate**

In each of the sections of the findings, we have tried to state limitations and assumptions. Those limitations and assumptions can lead to either exaggerated numbers or numbers that under-represent the problem. For example, above we explained that in the case of foreclosures, we made the assumption that people whose homes are being foreclosed live in the same proportion of poverty as does the general population. That is unlikely to be true; it is more likely that a person whose home is being foreclosed is more likely to be presently living in poverty than a person in the general population. On the other hand, we also acknowledge that the time period of the data we used is one of the worst foreclosure periods in our recent history, which would inflate our numbers.

Further, some of the limitations and assumptions are likely to impact our accuracy, but we cannot predict whether it increases or decreases our estimates. We have tried, as best we can, to make balancing assumptions, that is some likely to lead to lower outcomes and some likely to lead to higher outcomes. The goal is a reasonable estimate.

We are not going to repeat all of those here, but rather at a broader level examine the major thematic limitations of our approach, and how it might either inflate or deflate the estimate.

1.  Factors Leading to Overstatement

Throughout this study we assume that everyone facing the identified challenges needs legal assistance. This is certainly not always true. For example, some people with incapacities already have a guardianship. Some people facing domestic violence do not want an attorney, even if an attorney might be able to help them. Some landlords will voluntarily improve the conditions of a rental unit without the involvement of an attorney. In our analysis, we assume that some involvement by an attorney in each case will be beneficial.

2.  Factors Leading to Understatement

Throughout our study, we confronted large data gaps. As you can tell from the findings, thematically we were all frustrated by the lack of data. That systematic lack of data almost certainly across all areas produces an understatement.

For instance, the substandard housing analysis is significantly limited to only the 119 metropolitan areas included in the last three years of AHS surveys. This leaves out not only the entire rural areas of the United States, but it also leaves out many significant but slightly smaller cities and college towns across the country.

Another example is the lack of data on the economic circumstances of those facing foreclosure. The very fact that someone is facing foreclosure suggest that they are in challenging financial circumstances, but we have no data on that. Further, the National Institute of Justice includes stalking and psychological/emotional violence as forms of domestic violence (NIJ, 2007). Unfortunately, these forms of violence are beyond the scope of the NCVS and could not be captured in our analysis.

On balance, we fear that our numbers substantially understate the size of the problem.

C.  **Implications for Policy Makers**

1.  Justice

Hopefully, this analysis casts some useful light on the scope of need for legal aid in connection with the SDoH. The problems faced by those living in poverty are substantial and widespread throughout the United States. Indeed, we estimated that the total number of attorney hours to address just these 5 legal issues is over 34 million. To put that in perspective, the American Bar Association estimated that in 2018, the year for much of our data comes from, the total number of practicing attorneys in the United States was 1,338,678 (Weiss). Thus, to provide the needed hours, every single practicing attorney in the United States would need to contribute about 26 hours a year. While many lawyers do in fact contribute pro bono hours, they address the full range of legal needs that go well beyond just the five that we studied.

We found it interesting that our national numbers turned out to be similar for each of the issues in terms of the order of magnitude of attorney hours needed. Each of these needs is substantial. Collectively, the largest category is needs related to housing, whether affordability or suitability. That would suggest that investment in providing legal services expert in housing related issues is an area that needs to be examined. At the same time, more personal needs such as those of the cognitively disabled and those suffering domestic abuse that require a different skill set likewise could benefit from further investment.

2. Health

In order to achieve population health and control costs, stakeholders in healthcare industry are seeking solutions beyond medical interventions because the SDoH have a major impact on individual's health outcome. Since many of SDoH issues are legal problems such as quality of housing, domestic violence, etc, policy makers should look into a solution with multi-disciplinary approach. Investing in a coordinated approach involving both healthcare provider and community resources like legal service providers would seem fruitful. Connecting the patients with identified legal need with legal services is not new. However, funding for patients' legal interventions is challenging. Demonstrating the value and healthcare cost saving from medical-legal partnerships would be a potential way to engage healthcare payers to financially support patients' legal services.

## IX. Conclusions

In this analysis, we have looked only at need for legal services. We have not tried estimate supply. Ultimately, to assess the Justice Gap, we would need to calculate how much of this need is already being served. Unfortunately, we did not have the data to calculate that side of the equation.

While we did not know it in January 2020 when we started on this project, during the project it took on greater urgency as we began to appreciate how the data that we were looking at were almost certainly going to understate the need for these legal services in 2021. COVID has had a profound impact on American society, and as discussed throughout this paper, we began to understand how, for example, evictions and foreclosures are likely to spike up in 2021. We further noted that areas such as domestic abuse tend to spike during periods of extreme stress.

The needs here are urgent. While there are legal aid clinics, they almost certainly serve only a fraction of the 34 million total hours needed per year to address these legal needs. It is important to recognize further that when talking about the supply of legal services, that supply has to address much more than just the issues analyzed in this paper. Those living below poverty have a wide range of legal needs not studied in our analysis.

Nonetheless, we hope this partial analysis provides useful insight into the plight of those facing the Justice Gap.